\documentclass[12pt]{iopart}
\usepackage{indentfirst}
\usepackage{cite}
\usepackage{graphicx}
\usepackage{caption}
\usepackage{overpic}
\usepackage{float}
\usepackage{tabularx}
\usepackage{changepage}
\usepackage{threeparttable} 
\DeclareCaptionFormat{bold}{\textbf{#1#2}#3}
\usepackage{booktabs} 
 \usepackage{tikz}
\usetikzlibrary{positioning}

\begin{document}

\title[Identifying L-H transition in HL-2A
through deep learning]{Identifying L-H transition in HL-2A 
through deep learning}

\author{Meihuizi He$^1$,  Songfen Liu$^{1,\dagger}$, Fan Xia$^2$ , Zongyu Yang$^{2}$ and Wulyu Zhong$^2$}

\address{1 School of Physics, Nankai University, Weijin Road 94, Tianjin 300071, China}
\address{2 Southwestern Institute of Physics, PO Box 432, Chengdu 610041, China}
\ead{\mailto{$\dagger$ lsfnku@nankai.edu.cn}}
\vspace{10pt}

\begin{abstract}
During the operation of tokamak devices, addressing the thermal load issues caused by Edge Localized Modes (ELMs) eruption is crucial. Ideally, mitigation and suppression measures for ELMs should be promptly initiated as soon as the first low-to-high confinement (L-H) transition occurs, which necessitates the real-time monitoring and accurate identification of the L-H transition process. Motivated by this, and by recent deep learning boom, we propose a deep learning-based L-H transition identification algorithm on HL-2A tokamak. In this work, we have constructed a neural network comprising layers of Residual Long Short-Term Memory (LSTM) and Temporal Convolutional Network (TCN). Unlike previous work based on recognition for ELMs by slice, this method implements recognition on L-H transition process before the first ELMs crash. Therefore the mitigation techniques can be triggered in time to suppress the initial ELMs bursts. In order to further explain the effectiveness of the algorithm, we developed a series of evaluation indicators by shots, and the results show that this algorithm can provide necessary reference for the mitigation and suppression system.
\end{abstract}

%
\vspace{2pc}

\section{Introduction}
High-confinement Mode (H-mode) operation in tokamak devices, a critical operational mode for the International Thermonuclear Experimental Reactor (ITER) experiments, has been a central focus in fusion research\cite{ex2,ex3,ex4,ex5} since its discovery in 1982 \cite{ex1}.H-mode is frequently marked by ELMs bursts \cite{ex6}.These ELMs are accompanied with core high-temperature particles and energy release, subjecting first walls and divertors to intense thermal loads. This can potentially damage these critical components and posing a direct risk to the operational safety of the device. Given the ELMy-H mode's capability for effective energy confinement and impurity removal, precisely controlling ELMs amplitude offers the potential to promote the design of fusion device.

H-mode has been successfully replicated in a host of tokamaks under different configurations \cite{ex7,ex8,ex9}, but unfortunately, research on the physical mechanism of H-mode is still incomprehensive \cite{ex10}. Furthermore, the experimentally based power threshold scaling law for the L-H transition is highly uncertain, making a priori prediction of transition time during discharge challenging \cite{ex11}.

Experts can typically synthesize time series of multiple diagnostic signals after a discharge to distinguish various operating modes. However, when dealing with numerous discharges, human analysis becomes exceedingly complex and subjective, translating this human analysis into a real-time decision program is notably challenging. Recent studies \cite{ex12, ex13} indicate that deep learning methods are particularly adept at handling the high-dimensional data collected during experiments. Therefore, developing a data-driven method for automatically detecting certain events during discharge would be useful for tokamak experiments. For tokamaks operated in H mode, such a detector not only simplify the post-experimental data analysis but also, under ideal conditions, rapidly combine feedback on ELMs and plasma states into the real-time control system. This will facilitate the prompt implementation of measures such as Resonant Magnetic Perturbation (RMP) to safeguard the device's divertor and inner walls.

In experiments, the time range to apply measures such as RMP is typically pre-determined before the discharge process\cite{ex21,ex22}. However, the uncertainty of L-H transitions  during actual experiments often leads to either premature or delayed intervention of control measures. Recently, attempts have been made on devices like KSTAR\cite{ex18} to use Machine Learning (ML) classifier outputs for monitoring ELMs events, thereby leading to the triggering of RMP. A similar attempt was also made on the HL-2A, but due to algorithmic and device response delays, the initial few ELMs events are often overlooked\cite{ex15}. This is particularly critical if the initial ELMs are high in amplitude because even a single ELMs crash can impose significant thermal loads, which is strictly prohibited in ITER\cite{ex14}.This study implements a novel deep learning-based L-H transition recognition algorithm on the HL-2A tokamak, differing from previous slice algorithms\cite{ex15}. This method, combining Res-LSTM methods with TCN layers\cite{ex16,ex17}, can more effectively captures structured features in longer sequences. Supported by this approach, the algorithm shifts focus from transient ELMs bursts to the more extended L-H transition process, thus addressing the issue of missing initial ELMs events in current mitigation strategies.

This paper is organized as follows: Section 2.1 mainly introduces the channels of the input model and their definitions. Section 2.2 elaborates on the calibration process of the dataset with considerations for specific issues. Section 2.3 describes the model methods and network architecture used. Section 3 presents the training and testing results of the model and provides a detailed description of the evaluation methods employed. Section 4 delves into a thorough discussion of the potential flaws of the model. Finally, Section 5 summarizes the research of this paper and offers prospects for future research directions.

\section{Methods}
\subsection{Feature signal selection}

In previous studies, the deuterium spectral line of Balmer series $D_\alpha$ from the divertor and the line-averaged electron density $\bar{n_e}$ have typically selected as feature channels input in classifier \cite{ex18}. However, manually discerning the occurrence of L-H transition often presents a complicated challenge involving multi-dimensional channels. For example, distinguishing a decline of $D_\alpha$ spectral line due to changes in plasma configuration from the sharp decrease characteristic of L-H transition requires the integration of plasma horizontal and vertical position information.
For particularly contentious discharges, accurate analysis even requires higher-dimensional data, such as temperature and density profiles. Statistical analysis of discharge experiment data from the HL-2A device reveals the foundational operational range for high-confinement mode in the HL-2A tokamak: a plasma current ranging from 0.13 MA to 0.17 MA, stored energy between 20-50 kJ, and electron density greater than $1.5\times10^{19} m^{-3}$.For this work, we opted five diagnostic signals which experimental experts use to determine L-H transition after experiment and assembled a diagnostic dataset:

\begin{figure}[H]
            \centering
            \includegraphics[width=0.95\columnwidth,height=0.65\columnwidth]{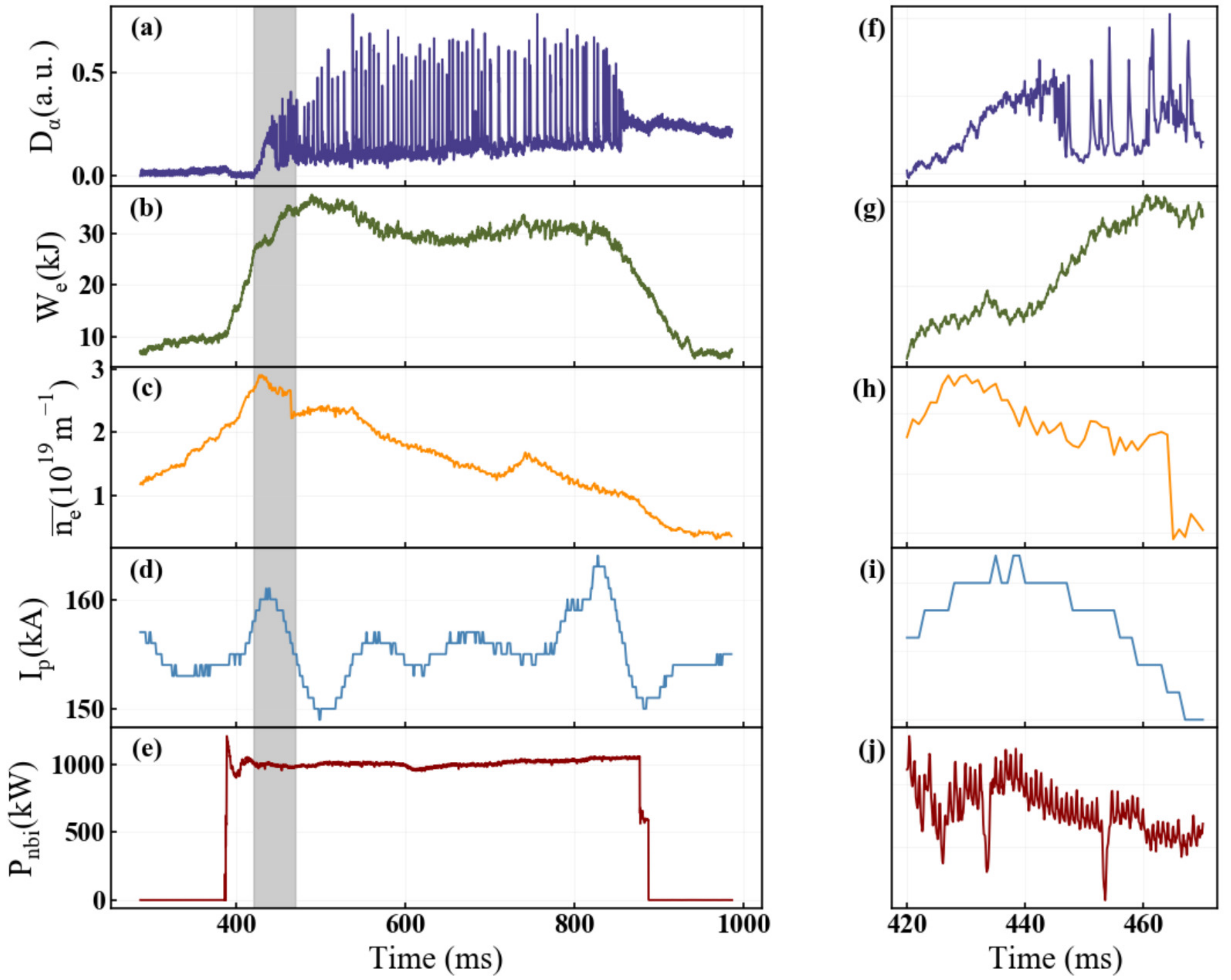}
            \caption{Schematic of the L-H transition identification diagnostic dataset, HL-2A Shot \#20000.
            Shown are (a) the $D_\alpha$ emission in the divertor region ;(b) plasma stored energy; (c) line-averaged electron density; (d) plasma current and (e) neutral beam injection power. (f)-(j) correspond to local magnification of the left channel at the L-H transition stage (grey region), respectively.}
\end{figure}
\begin{itemize}
\item \textbf{the deuterium spectral line of Balmer series from the divertor, $D_\alpha$:} By analyzing the $D_\alpha$ signal, different plasma mode transitions and ELMs burst that need to be suppressed can be easily observed. The main characteristic of the transition from L-mode to H-mode is a sudden drop in the $D_\alpha$ signal to the background level and then remains for a period. This is due to improved plasma confinement, resulting in a reduction of edge turbulence levels, which consequently decreases the number of particles entering the divertor.
However, as time goes by, the edge temperature and density gradients further increase, thereby inducing edge-localized instability ELMs. This is characterized by spiky oscillations in the $D_\alpha$ signal with very short relaxation times (approximately 2ms), corresponding to the large-scale expulsion of plasma stripping out and into the divertor during the ELMs burst.

\item \textbf{Neutral beam injection power, $P_{NBI}$:} One of the main auxiliary heating methods for HL-2A tokamak. Extensive research has shown that the transition to H-mode requires reaching a certain power threshold. Within the range of shots  studied in this paper for HL-2A, the typical H-mode only appears when the NBI is activated. Therefore, the dataset only considers shots with NBI signals, importing data from the start of NBI activation to 300ms after its deactivation. This operation effectively reduces the proportion of invalid data in the dataset, thereby accelerating the speed of algorithm training.

\item \textbf{Plasma stored energy, $W_E$:} During the L-H transition, the plasma's stored energy gradually increases. Statistical analysis of the H-mode diagnostic dataset indicates that H-mode typically does not occur when energy is below 20kJ.

\item \textbf{Line-averaged electron density, $n_e$:} This is obtained by measuring the chord-integrated electron density at the central channel (Z=-3.5cm) and dividing it by the measurement channel chord length. The transition threshold to H-mode is significantly correlated with the plasma density, and the critical power threshold increases significantly at a low density. According to current statistics, the transition power threshold corresponding to low densities is essentially unsatisfied on HL-2A.

\item \textbf{Plasma current, $I_P$:} This refers to the total plasma current. On HL-2A, auxiliary heating is typically only applied during the flat top phase when the current exceeds 50kA. Therefore, in considering the existence of L-H transition, moments when the absolute value of the current is below 50kA are ignored.
\end{itemize}

Figure 1 presents a localized magnification of these five channels during a typical H-mode discharge's transition. During the transition, the stored energy continuously rises, and upon completion of the transition, the $D_\alpha$ channel suddenly drops to a baseline level followed by periodic ELMs bursts. For subsequent analysis, it‘s necessary to resample these five signals, each with different sampling rates (i.e. the number of data points collected per second), to 10kHz. Given that a typical discharge event in the HL-2A device usually lasts for about 2 seconds, and the time scale for L-H transition is approximately 5-50ms, considering the potentially low proportion of valid data for H-mode occurrence, we have selected the data segment from 50ms before the start of NBI to 200ms after its complete shutdown as the training set. It should be noted that, in the actual testing phase, we still use the complete discharge sequence of a shot to ensure the completeness and accuracy of the experiment.

\subsection{Description of the dataset}
This study utilizes a dataset comprising 451 shots for training and 159 shots for testing. Due to the extensive span of discharge periods on HL-2A and continuous iterations of the diagnostic systems, the characteristics of the same signal channels have changed over time. For instance, in the later stages of HL-2A discharges, the $D_\alpha$ signal acquisition system increased its amplification factor, leading to overall higher values. However, the main feature of the $D_\alpha$ signal for L-H transition, characterized by initial oscillations followed by a rapid decrease, remains unchanged. With proper training of the identification algorithm, it is still possible to achieve accurate recognition, although the overall shape has not changed. To enhance the model's generalizability, 90\% of the data in the dataset is drawn from discharges between Shot 20000-25220, mixed with 10\% of typical discharge data after Shot 30000. After unifying the sampling rate of the signals and selecting our range of interest, each shot yields a sequence composed of approximately 7000 time points. Since this study employs supervised learning algorithms, manual labeling of the data is also required.

\begin{figure}[htpt]
            \centering
            \includegraphics[width=0.9\columnwidth,height=0.5\columnwidth]{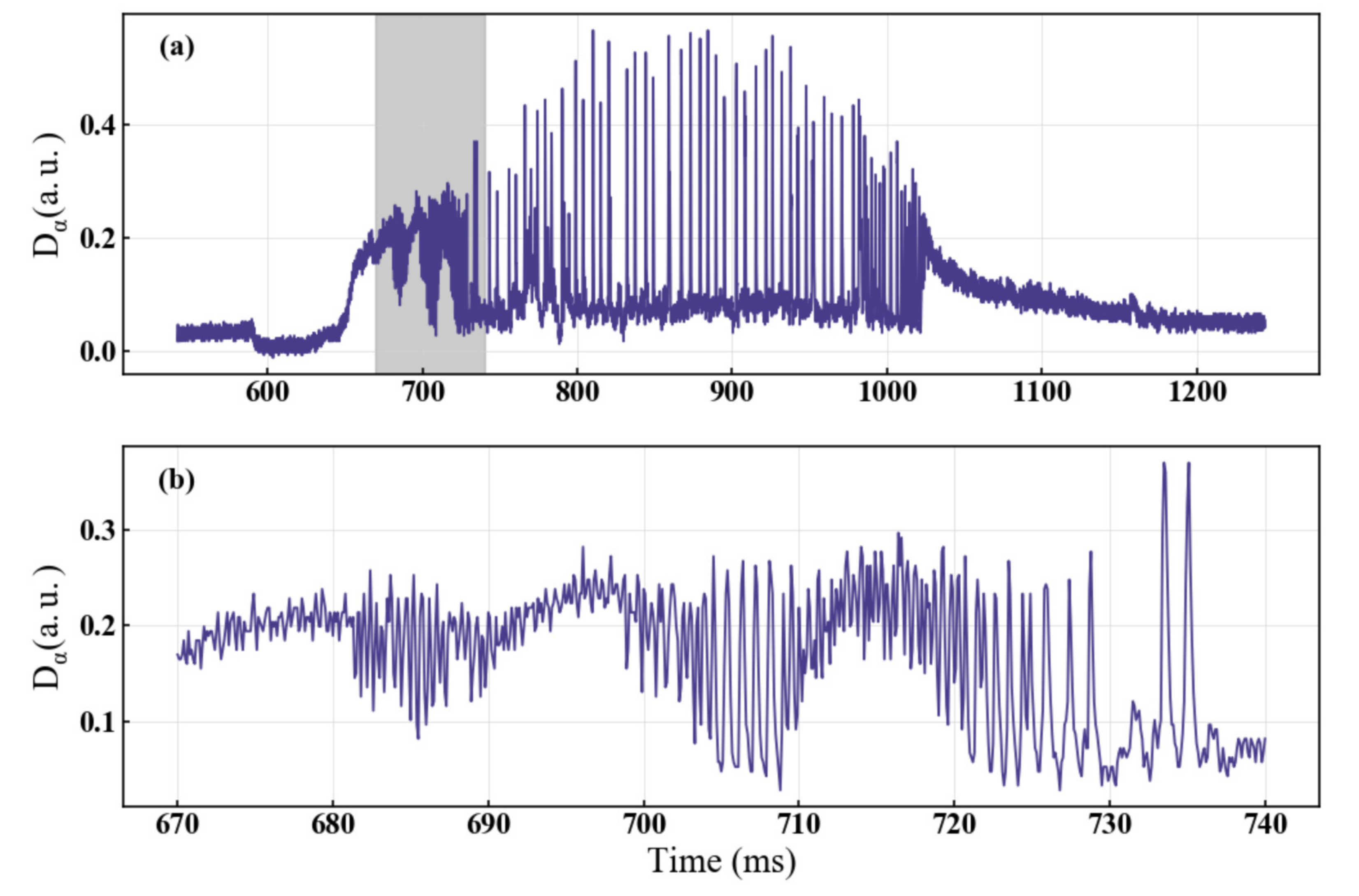}
            \caption{Shot \#20592. Figure (a) displays the intensity of the divertor $D_\alpha$ spectral line, while (b) presents a magnified view of the gray area.}
\end{figure}
Accurately pinpointing the exact moment of transition and ensuring consistency in the dataset labeling, even by the same person, is challenging due to several factors. Firstly, while the L-mode can be relatively easily identified, the transition to H-mode involves complex operating modes and plasma states such as dithering\cite{ex2} and Limit Cycle Oscillation (LCO), which add to the complexity of judgment. Secondly, the variability in data quality presents another challenge, some discharges exhibit rapid and clear transitions, while others show more scattered and irregular transition patterns. 
For example, a discharge might go through several 'near-transitions' --- showing some transitional behaviors but ultimately failing to enter a stable H-mode operation. As illustrated in Figure 2, multiple near-transitions are observed between 670-740ms in this discharge, but it didn't fully enter H-mode. Consequently, in the initial labeling phase, we tend to label the entire L-H transition process from the conclusion of L-mode to the point where the $D_\alpha$ signal drops and reaches a stable baseline. This approach results in significant differences in the characteristic behaviors and time spans of each shot, posing a considerable challenge for model training as it struggles to capture these varied and complex features.

The objective of this study is to identify the L-H transition rather than classify different transition phenomena. Meanwhile a pressing issue in the experiments is how to timely identify and alleviate the first two ELMs\cite{ex14}. In the application phase of the algorithm, the focus is on whether the signal pattern of a sharp decline in $D_\alpha$ followed by a plateau can be accurately captured. Considering these, three labels y={0, 1, 2}, are chosen to represent non-L-H transition intervals (label 0), $D_\alpha$ decline period (label 1), and the intermediate phase (label 2). These labels are intended to facilitate correct identification of the L-H transition process in algorithm training. Figure 3 shows the segmentation method for a sample discharge, and the specific method for setting segments is as follows:

\begin{enumerate}

\item Label 0: Assigned to time points that do not involve the L-H mode transition process. This includes stable periods before the transition and after the transition is completed, indicating that during these time intervals, the signals do not exhibit characteristics of transition.

\item Label 1: Specifically used to denote data during the L-H transition process. More precisely, it marks a narrow window around the completion time of the transition — from 5ms before to 5ms after the transition. This interval typically includes the $D_\alpha$ decline period and the first two ELMs appear. In subsequent algorithm training, we expect the algorithm to output values close to 1 in this interval, thereby identifying the moment of L-H transition.

\item Label 2: Applied as the intermediate phase of L-H transition that cannot be accurately identified, used in two scenarios. Firstly, it marks data points 5ms before the transition starts and other time points during the transition process. Secondly, it is also used for a specific post-transition interval— from 5ms to 10ms after the completion of the transition. The rationale for Label 2 is to train the algorithm to process transitional period data properly, avoiding overfitting issues from hard labeling uncertain data. In this case, the model's classification of 0, 1, or 2 for such data is considered correct, ensuring the model's generalizability.
\end{enumerate}

\begin{figure}[htpt]
            \centering
            \includegraphics[width=0.95\columnwidth,height=0.5\columnwidth]{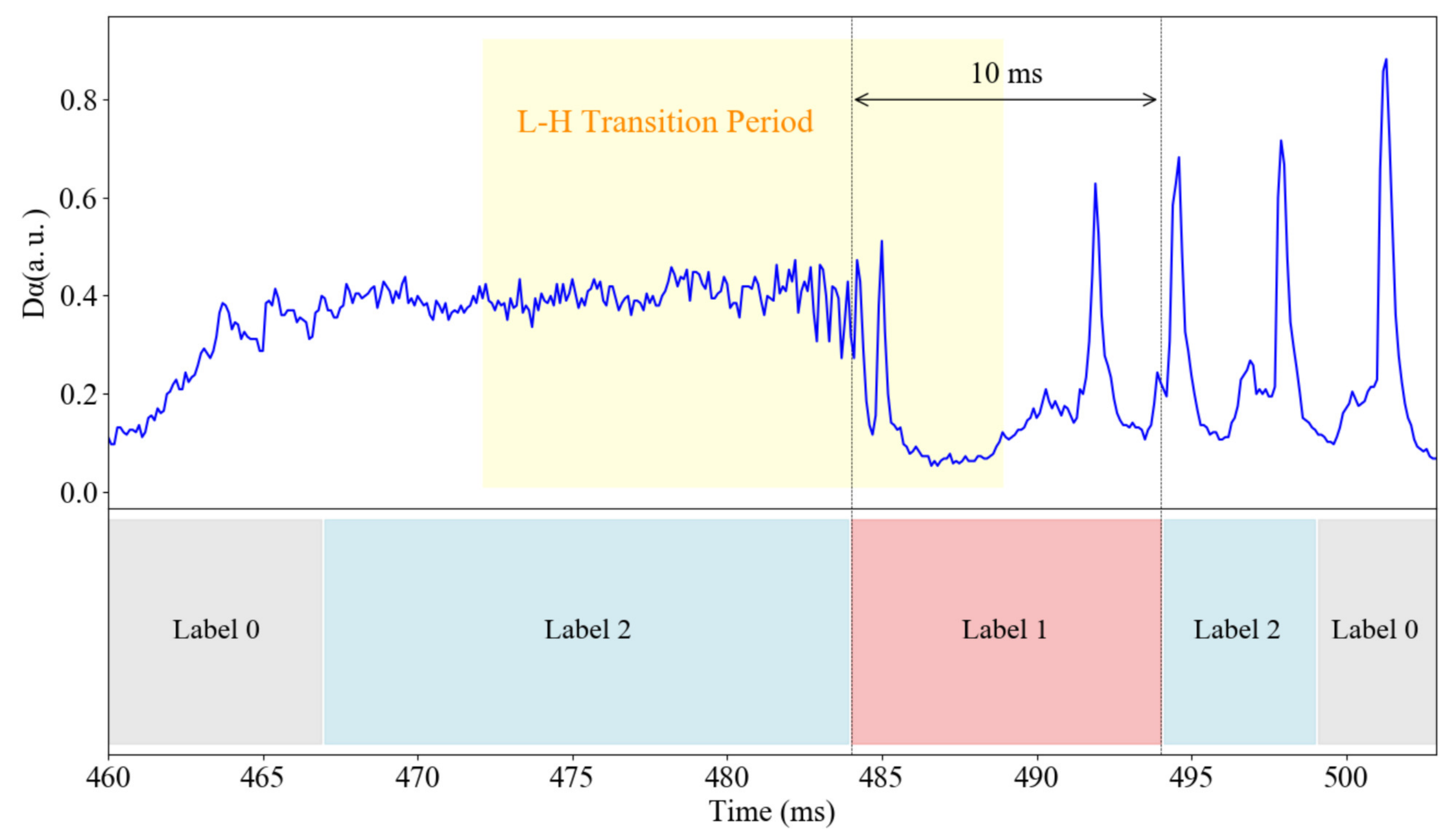}
            \caption{Schematic diagram of conversion label assignments for Shot \#26592. The yellow area in the upper subplot shows the manually labelled L-H transition time interval.}
\end{figure}
After completing the data labeling, normalization of the data is necessary for adapting to a deep learning environment. The normalization process retains amplitude information to ensure reproducibility during the model inference stage. Therefore, it involves subtracting the global mean of the corresponding channel across the entire training set, followed by division by its global standard deviation. Additionally, to ensure consistency between the test set and the training set, the mean and standard deviation calculated from the training set are stored and used to normalize the test set data.

\subsection{Model design}
In this study, a combination of Res-LSTM networks and TCN is utilized for deep analysis and processing of data. This joint network structure ensures computational efficiency and extracts rich spatial-temporal information while handling data with complex dynamic characteristics. Consequently, during the L-H transition, it accurately captures subtle state changes and the dynamic evolution of related physical processes, providing powerful data support for understanding and recognizing key mechanisms in the transition process. The specific model structure is shown in Figure 4:

\begin{figure}[H]
    \centering
    \includegraphics[width=\columnwidth,height=0.5\columnwidth]{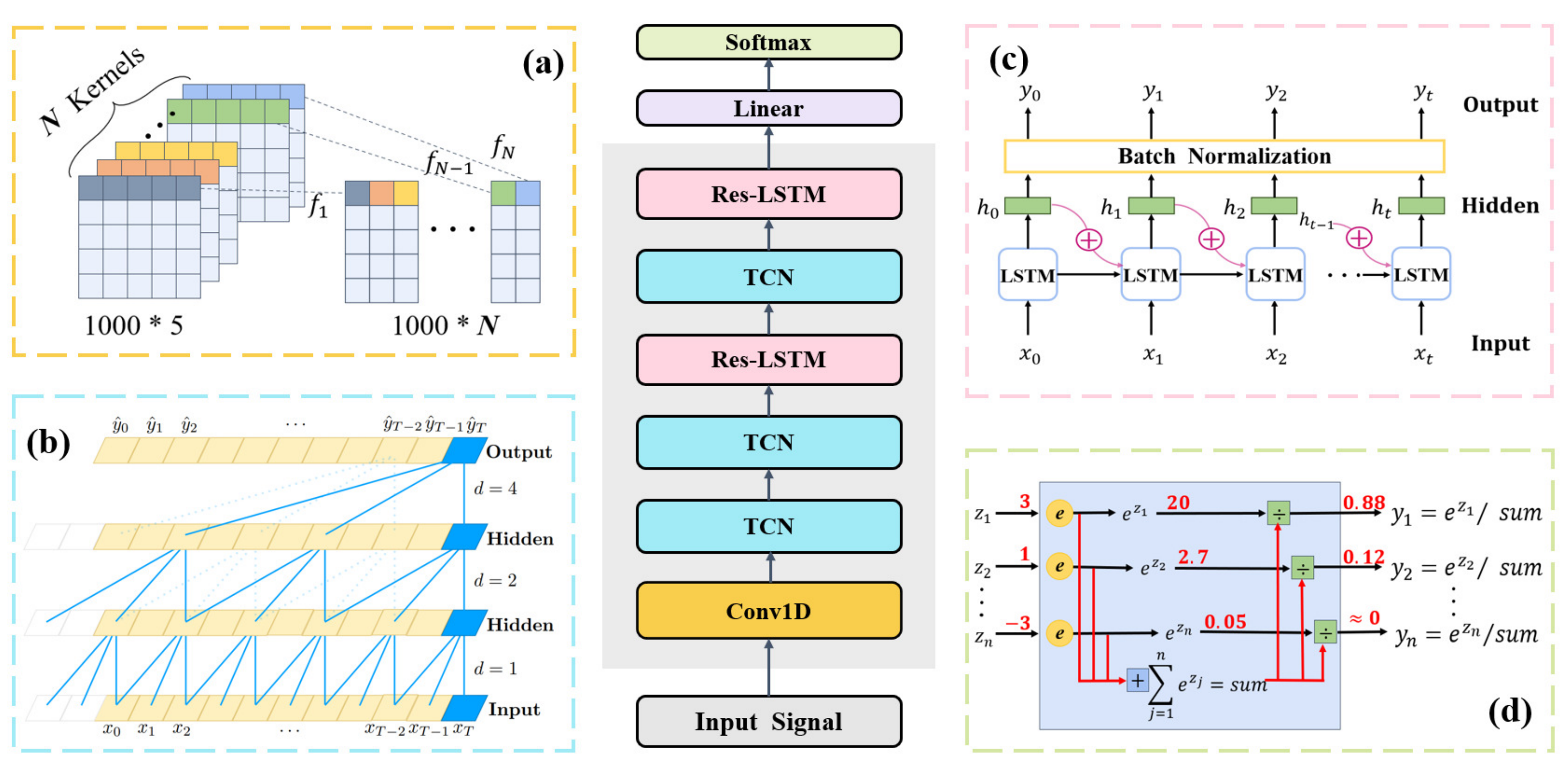}
    \caption{Schematic of the neural network model used in this study.
    (a) Conv1D layer, employing multiple convolutional kernels to extract features from time series data.
    (b) TCN layer, using causal and dilated convolutions to capture long-span behaviors without information leakage.
    (c) Res-LSTM layer, enhancing gradient flow and information processing ability through gated units and residual connections.
    (d) Softmax layer, transforming processed features into a probability distribution for state classification.}
    \label{fig:neural_network_model}
\end{figure}

\begin{itemize}
\item {\bf Input layer:} The input receives a series of matrices preprocessed and reshaped into 1000×5 format. Here, '1000' represents the number of time steps, each comprising 5 feature dimensions. This ensures the capture of critical L-H transition dynamics within a 100ms window, which slides across the sequence with a fixed step length of 5ms.  During the testing phase, a non-overlapping slicing strategy is applied to ensure continuity and completeness in processing individual discharge data.

\item{\bf Convolutional layer (Conv1D):} As the initial stage of feature extraction shown in Figure 4(a), this layer employs one-dimensional convolutional kernels that slide along the time axis of the input sequence. Parameters of each kernel are determined by dot product with input data, generating new multidimensional feature maps. This process is designed to extract local features efficiently and provides comprehensive spatiotemporal information for subsequent layers.

\item{\bf Temporal Convolutional Network (TCN):} Figure 4(b) highlights the essential features of the TCN layer. The blue arrows depict the utilization of dilated convolution, a technique that introduces gaps into the standard convolutional map to expand the receptive field. This technique allows each node access a broader range of historical data, significantly enhancing its receptive field without including future information.

\item{\bf Residual Long Short-Term Memory Network (Res-LSTM):} As depicted in Figure 4(c), the Res-LSTM structure processes temporal information at each time step is first processed by LSTM units and then outputted through residual connection. This substantially improves the flow of information in the deeper network layers. Within the LSTM units, a coordinated work of forget gates, input gates, and output gates carefully regulates the information flow\cite{ex16}. This allows the network to update its memory based on the current input and past states in a targeted way. 
By adding residual connections, the risk of gradient vanishing in deep networks is reduced, and the network's learning is simplified by reducing its reliance on identity mapping. While focusing on the differences between inputs and outputs, the model becomes more efficient in training and improves its overall performance.

\item{\bf Linear layer:} The linear layer acts as a bridge linking complex time series features with the final classification decision, mapping the output of the LSTM layer to the category space through linear transformations.

\item{\bf Softmax Layer:} As illustrated in Figure 4(d), the Softmax layer functions in the model's final stage, converting the linear layer's output into a probability distribution. It processes the scores for each category using the Softmax function, thereby generating normalized probabilities. In the task of recognizing L-H transition, this layer outputs a probability vector that represents the predicted probabilities for three different plasma states, providing a basis for the final classification decision.
\end{itemize}

This study uses an improvement over the traditional Adam algorithm, the AdamW optimizer\cite{ex19}. AdamW enhances regularization by separating weight decay from the gradient update process, thereby improving the model's generalization ability.  The learning rate and weights are dynamically adjusted, decreasing with each training epoch. This dynamic adjustment refines the optimization process in the latter stages of training and effectively prevents overfitting. Furthermore, the study incorporates an early stopping mechanism as a part of the training callback function. This mechanism halts the training process if no improvement in the AUC value is observed on the validation set over 15 continuous epochs.Upon halting, the model automatically reverts to the previously recorded best weights. This strategy is designed to ensure that the model achieves optimal performance in its subsequent applications.

The optimization of hyperparameters for training the model is an iterative process that comprehensively considers computational efficiency and model performance. This process encompasses a range of parameters, from basic ones like learning rate and training epochs to specific configurations of network architectures, such as the LSTM and TCN layers. We evaluated the effectiveness of various parameter combinations through a series of experiments. The optimal hyperparameters, as identified and listed in Table 1, were applied in subsequent testing and assessments, ensuring the model's superior performance.

\begin{table}[htbp]
\captionsetup{singlelinecheck=off, justification=raggedright}
\caption{Hyperparameter settings and explanations with their best values.}
\label{tab:hyperparameters}
\begin{tabular}{@{}lll}
\toprule
Hyperparameter & Explanation & Best Value \\
\midrule
$\eta$ (learning rate) & Initial learning rate for AdamW optimizer & 1e-3 \\
$\lambda$ (weight decay) & Initial weight decay for AdamW optimizer & 1e-4 \\
Batch\_size & Batch size for training and validation & 150 \\
Epochs & Maximum number of epochs to run training & 100 \\
GaussianNoise & Standard deviation of the noise distribution & 0.01 \\
RandomZoom & Zoom range for height and width & 0.1 \\
Conv1D (filters, kernel, strides) & Conv1D layer configuration & (32, 1, 1) \\
Dropout (TCN) & Dropout probability for TCN layers & 0, 0.2, 0.5 \\
LSTM units & Number of units in LSTM layers & 32 \\
Filters (TCN) & Number of filters in TCN convolutions & 64, 32, 16 \\
Kernel\_size (TCN) & Kernel size in TCN convolutions & 3 \\
Dilation\_rate (TCN) & Dilation rate in TCN convolutions & 2, 2, 4 \\
TimeDistributed Dense units & Number of units in TimeDistributed layer & 64 \\
Activation function (Dense) & Activation function for Dense layer & RELU \\
\bottomrule
\end{tabular}
\end{table}

\section{Model Results Evaluation}
\subsection{Evaluation of Algorithm Performance by Slice}
The classification of labels adopted in this study reveals a significant sample imbalance issue: the L-H transition process (including Label 1 and Label 2) accounts for less than 5\% of the total dataset in terms of the temporal scale. Considering the uncertainty associated with the phase indicated by Label 2, it was assigned a minimal weight. This approach ensures that the model leverages this segment of data during training without biasing the final predictions towards Category 2. Although the output is a three-class problem, the problem fundamentally remains a binary classification of the L-H transition period. Consequently, to address the discussed sample imbalance, the weights for the samples were set to $\omega_0 : \omega_1:\omega_2 =15:400:3$. This weighting scheme not only balances the class distribution within the dataset but also ensures full use of the positive samples during model training, while maintaining a certain focus on the ambiguous regions.

In evaluating the classifier model's performance on the test set, the intermediate Label 2 is recognized for its potential to be reasonably classified as either positive (Label 1) or negative (Label 0) during prediction. The assessment strategy for samples initially classified as Label 2 adheres to the following principle: if the model predicts these samples as positive, they are reclassified as Label 1 for evaluation purposes; similarly, if predicted as negative, they are treated as Label 0. This approach transforms the evaluation into traditional binary classification, ensuring both consistency and fairness. As previously mentioned, this treatment avoids assigning definitive labels to ambiguous data during training and enables the model to accurately identify the true nature of such data.

The assessment of model performance is critically dependent on the alignment between predicted and true labels. To that end, we build the L-H transition Confusion Matrix, which defines several variables: True Positive (TP) represents the count of positive samples accurately predicted as positive, while True Negative (TN) denotes the count of negative samples accurately predicted as negative. Conversely, False Negative (FN) refers to positive samples incorrectly predicted as negative, and False Positive (FP) indicates negative samples mistakenly classified as positive. 

By using these metrics to indicate a comprehensive assessment of the model's accuracy and reliability in recognizing L-H transition, one can then compute four critical metrics: True Positive Rate (TPR), True Negative Rate (TNR), False Positive Rate (FPR), and False Negative Rate (FNR). These metrics, whose calculations are delineated below (\ref{eq1}) and (\ref{eq2}), offer a quantitative evaluation of the model's performance at varying detection thresholds:
\begin{eqnarray}
\rm{T P R} & =\frac{\rm{T P}}{\rm{T P+F N}}   , \quad \rm{T N R} & =\frac{\rm{T N}}{\rm{T N+F P}}\label{eq1}\\
\rm{F P R} & =\frac{\rm{F P}}{\rm{F P+T N}}  , \quad \rm{F N R} & =\frac{\rm{F N}}{\rm{T P+F N}} \label{eq2}
\end{eqnarray}

Table 2 presents the model's above values at the optimal threshold when applied to the test set.

\begin{table}[htbp]
\captionsetup{singlelinecheck=off, justification=raggedright}
\begin{adjustwidth}{10mm}{} 
\caption{Four classification evaluation metrics for test set}
\label{tab:model_accuracy}
\small 
\begin{tabular}{@{}lllll@{}}
\toprule
  & TPR & TNR& FPR &FNR\\
\midrule
Test & 0.933 & 0.953& 0.067& 0.047 \\
\bottomrule
\end{tabular}
\end{adjustwidth}
\end{table}
In evaluating models for imbalanced class distribution datasets, ROC (Receiver Operating Characteristic) curve analysis is utilized, involving adjustments to various detection thresholds. These thresholds demarcate the classification boundary in predictive outputs: predictions surpassing the threshold are deemed positive, and those below, negative. In Figure 5, the ROC curve displays the TPR on the vertical axis and the FPR on the horizontal axis. Here, TPR indicates the model’s ability to correctly identify L-H transition (akin to a success rate), while FPR measures the proportion of non-L-H transition incorrectly identified as such (akin to a false alarm rate). Ideally, the model’s ROC curve should approximate the top left corner, denoting high success rates even at reduced false alarm rates. Figure 5 also marks TPR and TNR values at the optimal classification threshold. The Area Under the Curve (AUC) indicates the model's overall performance across various decision thresholds. The AUC value closer to 1 denotes superior model classification performance.
\vspace{-0.5em}
\begin{figure}[htpt]
    \centering  \includegraphics[width=0.6\columnwidth,height=0.45\columnwidth]{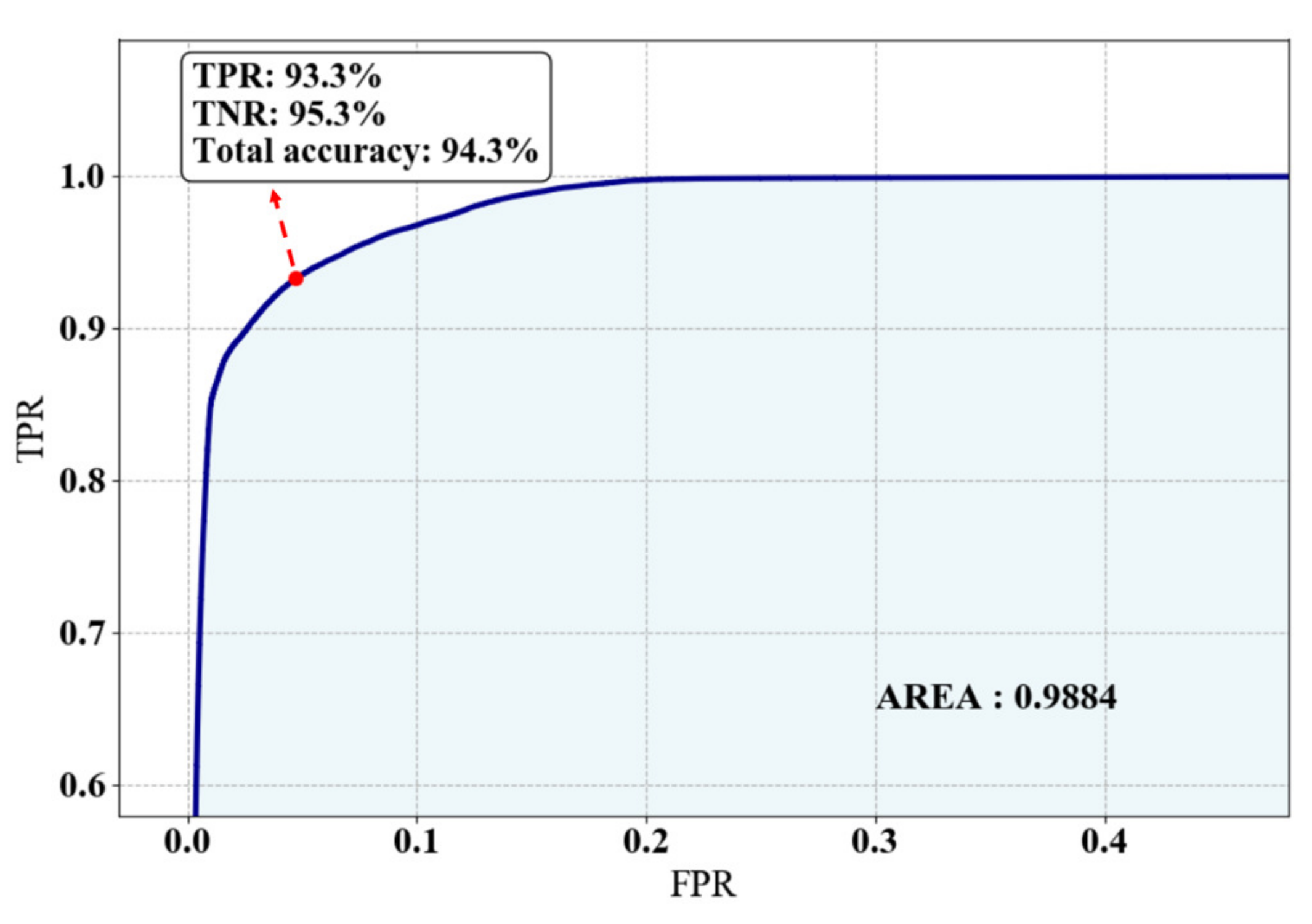}
    \caption{ROC Curve of the Model for test set with an AUC Value of 0.9884.}
    \label{fig:auc}
\end{figure}
\subsection{Evaluation of Algorithm Performance by Shot}

Figures 6-7 display the model's prediction results. In these graphs, the background color of the upper subgraph represents the output state of the neural network, while the background color in the lower subgraph indicates the real given label.

\begin{figure}[htpt]
    \centering
    \includegraphics[width=0.95\columnwidth,height=0.37\columnwidth]{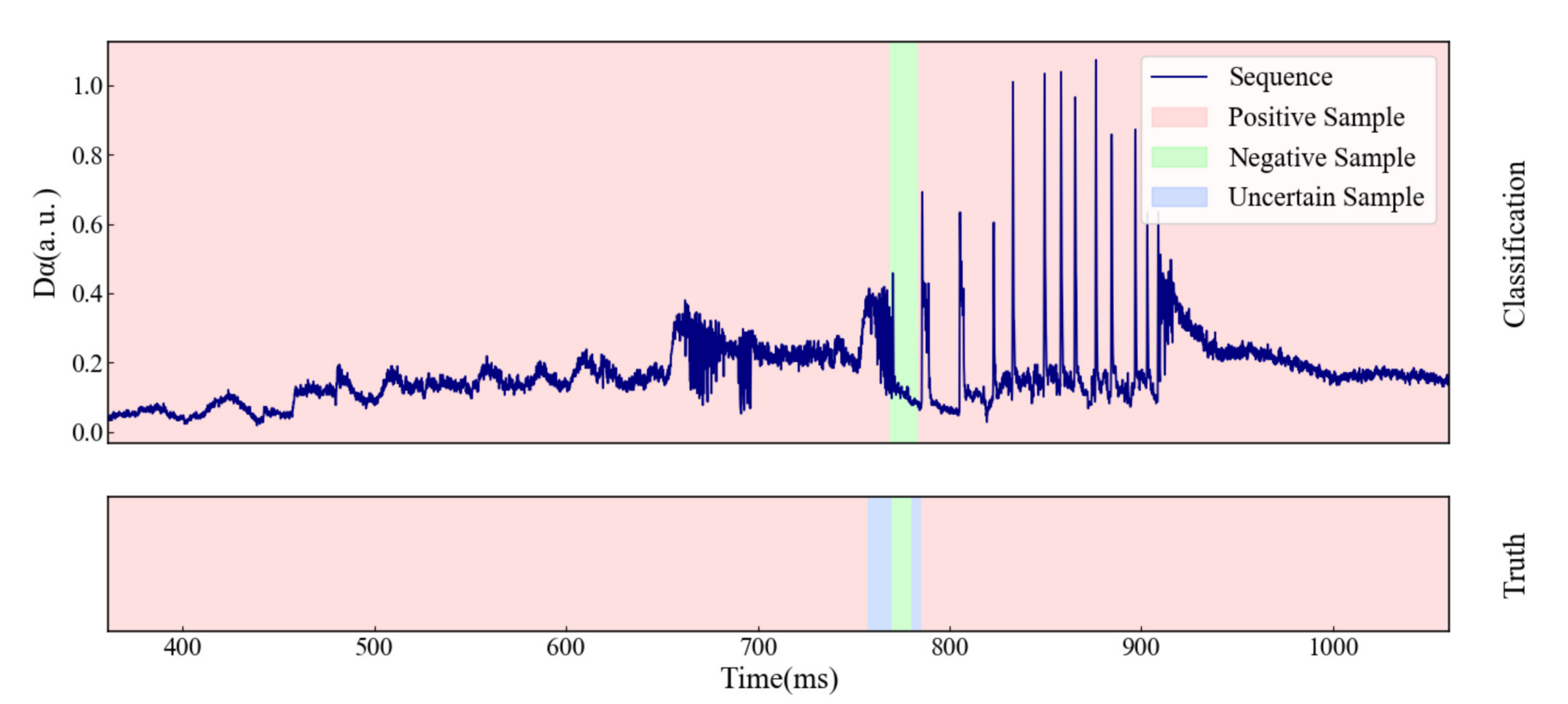}
    \caption{Prediction Results for Discharge Shot \#21954. The model effectively recognizes the L-H transition periods marked by a notable drop in $D_\alpha$ signal.}
    \label{fig:neural_network_model}
\end{figure}
\begin{figure}[H]
    \centering
    \includegraphics[width=0.95\columnwidth,height=0.37\columnwidth]{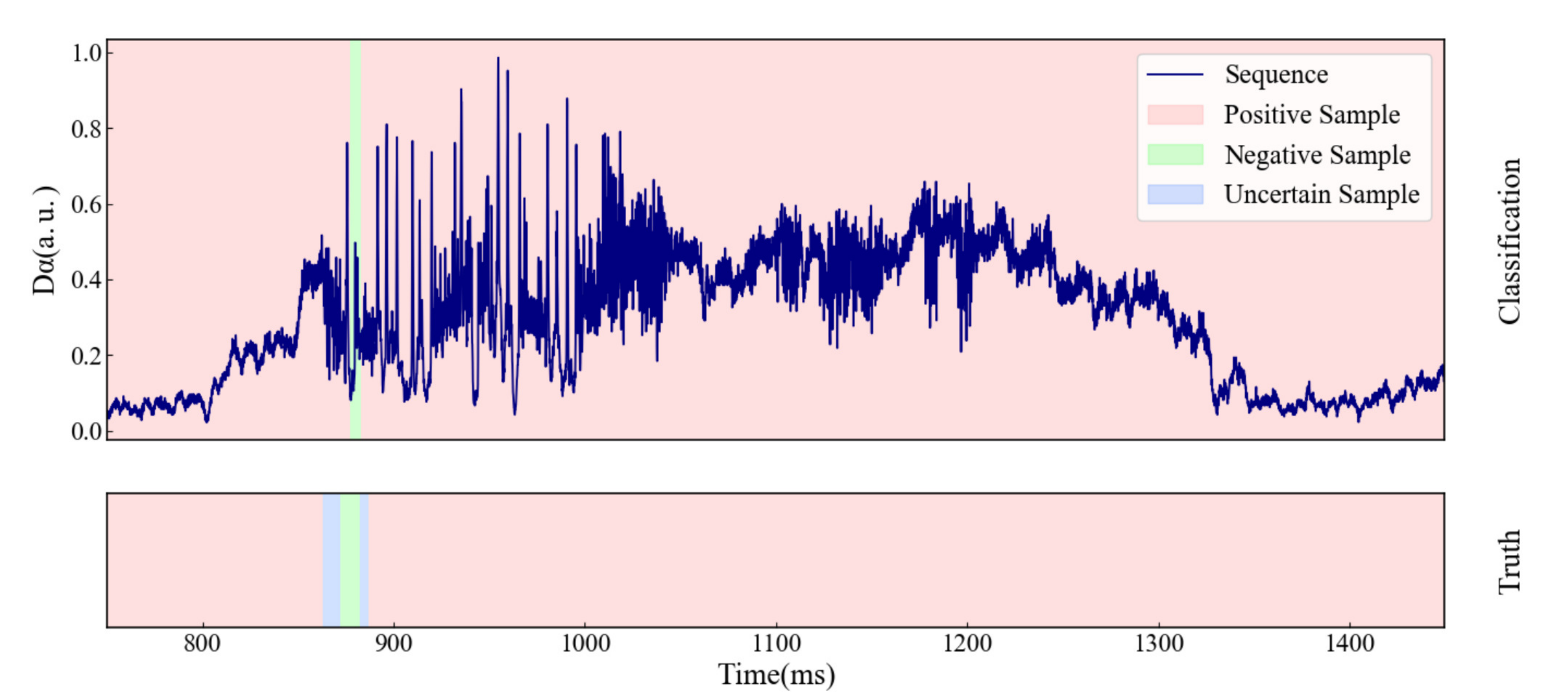}
    \caption{Prediction Results for Discharge Shot \#34456. The shot exhibits a complex transition, which briefly entering and exiting H-mode. The model also demonstrates accurate predictions in this scenario, as shown in the graph.}
    \label{fig:neural_network_model}
\end{figure}
\vspace{-1em}
\begin{figure}[H]
    \centering
    \includegraphics[width=0.65\columnwidth,height=0.8\columnwidth]{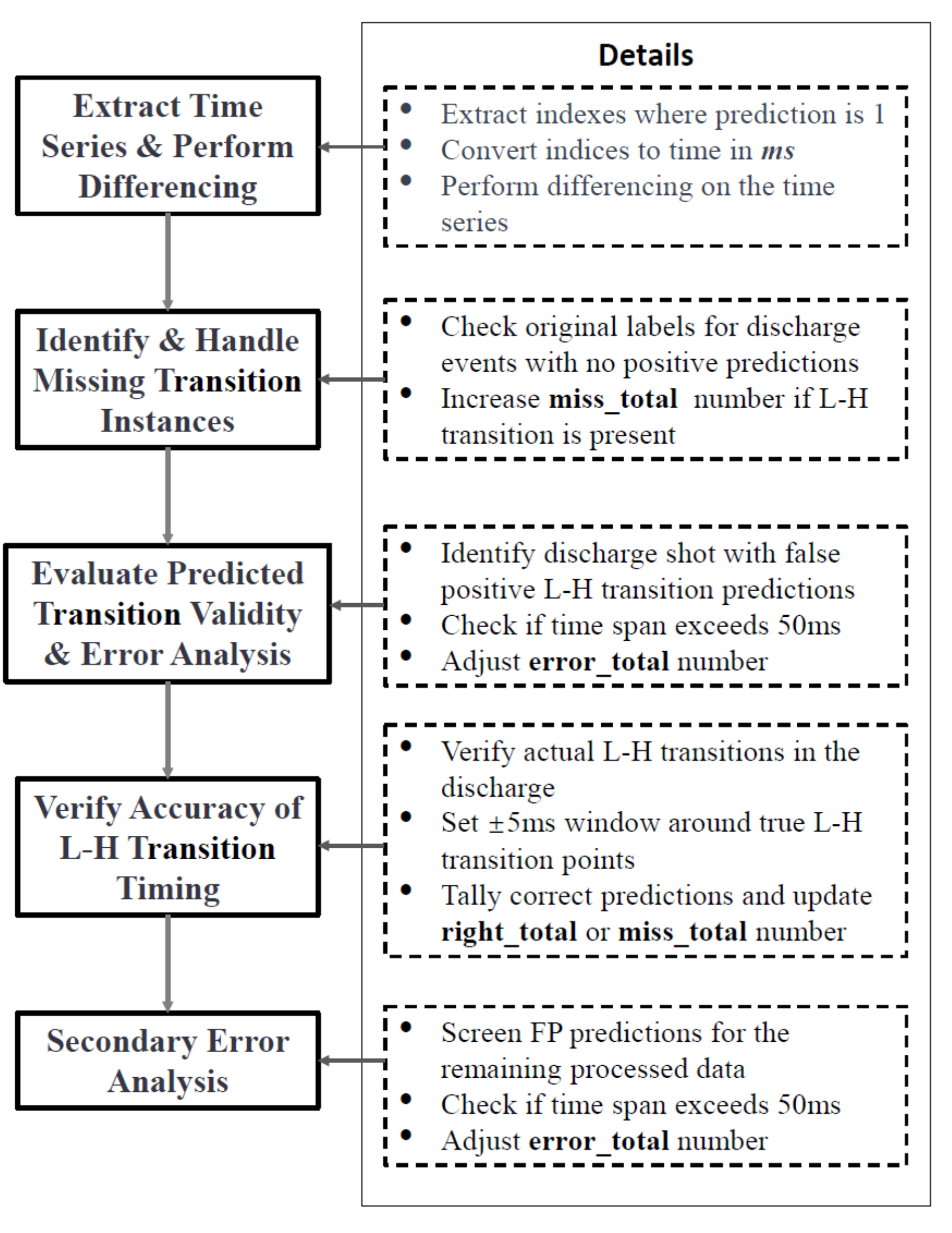}
    \caption{Flowchart of Shot-Based Performance Metrics for Algorithm Evaluation.}
    \label{fig:neural_network_model}
\end{figure}
Traditional ELMs recognition algorithms evaluate accuracy through statistics derived from smoothed temporal slice outputs. Yet, this approach is heavily influenced by the slicing methodology, making it difficult to compare results across different studies and often failing to reflect the true efficacy in experimental applications. Considering our study's time series model, which produces outputs at each time point, we introduced a series of evaluation metrics based on individual shots. These encompass the total number of valid test shots \textbf{(shot\_total)}, L-H transitions \textbf{(LH\_total)}, correct identifications \textbf{(right\_total)}, total number of misidentifications \textbf{(error\_total)}, and missed transitions \textbf{(miss\_total)}. Figure 8 presents the process flow for obtaining the above evaluative metrics, offering a detailed view of the algorithm’s performance assessment process.

Upon detailed analysis, within the test set comprising 161 discharge shots, the algorithm correctly detected 67 of the 85 L-H transition events, which missing 18 events. In addition, there were 16 false identifications of L-H transitions. The shot-based accuracy rate, while lower than the slice-based statistical outcome, offers a more direct representation of the algorithm’s practical efficacy in experimental application stages.
\section{Summary and future works}
In this study, we have successfully developed and validated a deep learning model that integrates residual LSTM and TCN networks, tailored for the precise prediction of L-H mode transitions in the HL-2A device. Demonstrating exemplary performance, the model achieved  an AUC of 98.8\% in the test dataset(\ref{fig:auc}), underscoring its effectiveness in addressing this complex task.

To ensure robustness and reliability in the predictive outcomes, a specialized statistical method for shot-based analysis was implemented. This method involved analyzing key metrics, namely the total number of valid test shots, the total L-H transitions, the correctly identified transitions, misidentifications, and missed transitions. The assessment of these metrics in the test dataset, which included 161 shots, revealed that the model accurately detected 67 of the 85 L-H transition events, missing 18 events. Furthermore, the model incorrectly identified 16 non-transition events as L-H transitions. This approach, grounded in rigorous statistical analysis, provides a clear and systematic evaluation of the model's performance in a real-world application context.

Above results indicate that the model has the potential for promising applications. In particular, by instantly and accurately obtaining the feedforward information of the L-H transition, the present model is able to provide timely information to support the mitigation measures for triggering the ELMs. This helps to take control strategies in advance to adjust the amplitude of the ELMs, thus protecting the device from damage caused by plasma particle fierce collision and high thermal loads.

However, the results of the statistical indicators show that the model still suffers from several misidentifications. This can be attributed to two main factors: firstly, the number of negative samples far exceeds the number of positive samples, which is still less than optimal even though we have taken into account the effect of weights. Secondly, the model has some challenges in distinguishing between 'near-transitions' and real transitions, because the current algorithm still relies too much on the $D_\alpha$ signal for its judgment, and from the perspective of the L-H transitions, the two do behave very similarly in the currently selected channel. For future research, as the temperature profile and density profile databases continue to improve, we plan to develop a time-series-based model of the H-mode power threshold scaling. This may help to improve the accuracy of recognizing L-H transitions in contextually specific.

The model introduced in this paper, incorporating a combined architecture of residual LSTM and TCN networks, has exhibited outstanding capabilities. This establishes an essential foundation for its applicability and efficiency in areas such as complex data analysis, real-time surveillance, and intelligent control in fusion plasma studies. Looking ahead, further studies and enhancements should concentrate on the model's adaptability across different physical conditions and delve into a broader range of application scenarios and actual strategies.

\ack{This study was supported by National Natural Science Foundation of China under grant No. 12275142 \& No. U21A20440. The authors would like to thank the HL-2A collaboration group and the CODIS team for their invaluable support.}
\section*{References}
\bibliographystyle{iopart-num}
\bibliography{iopart-num}

\end{document}